\documentclass[10pt]{amsart}
\usepackage{amssymb}
\usepackage{amsmath,amssymb,amsfonts,amsthm,
latexsym, amscd, amsfonts, epsfig, eepic,epic,psfrag}
\usepackage{mathrsfs}
\usepackage{color}
\usepackage{eucal}%    caligraphic-euler fonts: \mathcal{ }
\usepackage{eufrak}%   frak-euler        fonts: \mathfrak{ }
\usepackage[all]{xypic}
\usepackage{xspace}

\theoremstyle{plain}
\newtheorem{theorem}{Theorem}

\newtheorem{lemma}{Lemma}

\theoremstyle{definition}
\newtheorem{definition}{Definition}

\theoremstyle{example}

\newtheorem{remark}{Remark}
\theoremstyle{remark}

\numberwithin{equation}{section}

\begin{document}

%%%
%%%
%%%%%%%%%%%%%%%%%%%%%%%%%%%%%%%%%%%%%%%%%%%%%%%%%%%%%%%%%%%%%%%%%%%%%%%%%%
%%
%%%
\title[RNA-LEGO: Combinatorial Design of Pseudoknot RNA]
      {RNA-LEGO: Combinatorial Design of Pseudoknot RNA}
\author{Emma Y. Jin and Christian M. Reidys$^{\,\star}$}
\address{Center for Combinatorics, LPMC-TJKLC %XXX%
           \\
         Nankai University  \\
         Tianjin 300071\\
         P.R.~China\\
         Phone: *86-22-2350-6800\\
         Fax:   *86-22-2350-9272}
\email{reidys@nankai.edu.cn}%XXXX
\thanks{}
\date{November, 2007}
\begin{abstract}
In this paper we enumerate $k$-noncrossing RNA pseudoknot structures with
given minimum stack-length. We show that the numbers of $k$-noncrossing
structures without isolated base pairs are significantly smaller than the
number of all $k$-noncrossing structures. In particular we prove that the
number of $3$- and $4$-noncrossing RNA structures with stack-length
$\ge 2$ is for large $n$ given by
$311.2470 \, \frac{4!}{n(n-1)\dots(n-4)}2.5881^n$
and $1.217\cdot 10^{7} n^{-\frac{21}{2}} 3.0382^n$,
respectively.
We furthermore show that for $k$-noncrossing RNA structures the drop
in exponential growth rates between the number of all structures and
the number of all structures with stack-size $\ge 2$ increases significantly.
Our results are
of importance for prediction algorithms for pseudoknot-RNA and
provide evidence that there exist neutral networks of RNA pseudoknot
structures.
\end{abstract}
\maketitle
{{\small
%\tableofcontents
}}

%%%
%%%
%%%%%%%%%%%%%%%%%%%%%%%%%%%%%%%%%%%%%%%%%%%%%%%%%%%%%%%%%%%%%%%%%%%%%%%%
%%%
%%%

\section{Introduction}\label{S:intro}

\begin{figure}[ht]
\centerline{%
\epsfig{file=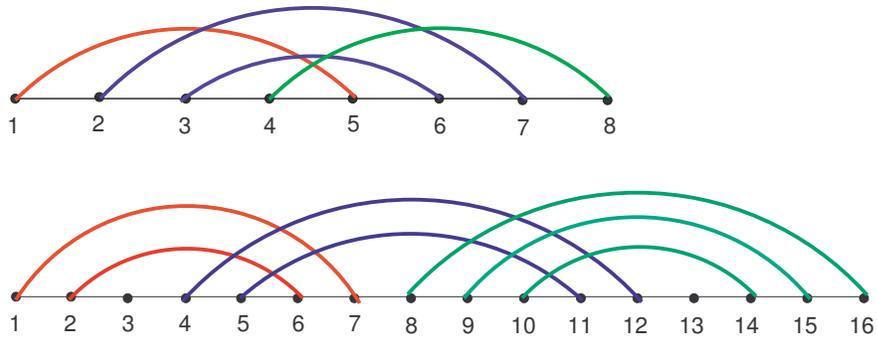,width=0.8\textwidth}\hskip15pt
 }
\caption{\small $k$-noncrossing diagrams: in the the upper diagram
the arcs red/blue/green mutually cross, the arc with minimum length
$3$ is $(3,6)$ and the arc $(1,5)$ is isolated. Hence this is a
$4$-noncrossing, $\lambda=3$, $\sigma=1$ diagram without isolated
vertices. Analogously, below we have a $3$-noncrossing (no red/green
cross), $\lambda=4$, $\sigma=2$ diagram with isolated vertices
$3,13$. } \label{F:lego1}
\end{figure}

%%%
%%%%%%%%%%%%%%%%%%%%%%%%%%%%%%%%%%%%%%%%%%%%%%%%%%%%%%%%%%%%%%%%%%%%%%%%
%%%
An RNA structure is the helical configuration of an RNA sequence,
described by its primary sequence of nucleotides {\bf A}, {\bf G},
{\bf U} and {\bf C} together with the Watson-Crick ({\bf A-U}, {\bf
G-C}) and ({\bf U-G}) base pairing rules. Subject to these single
stranded RNA form helical structures. Since the function of many RNA
sequences is oftentimes tantamount to their structures, it
is of central importance to understand RNA structure in the context
of studying the function of biological RNA, as well as in the design
process of artificial RNA. In the following we use a coarse grained
notion of structure by concentrating on the pairs of nucleotide
positions corresponding to the chemical bonds and ignoring any
spatial embedding. There are several ways to represent these RNA
structures \cite{Schuster:98a,Waterman:94a}. We choose the diagram
representation \cite{Stadler:99} which is particularly well suited
for displaying the crossings of the Watson-Crick base pairs. A
diagram is a labeled graph over the vertex set $[n]=\{1, \dots, n\}$
with degree $\le 1$, represented by drawing its vertices $1,\dots,n$
in a horizontal line and its arcs $(i,j)$, where $i<j$, in the upper
half-plane. The vertices and arcs correspond to nucleotides and
Watson-Crick ({\bf A-U}, {\bf G-C}) and ({\bf U-G}) base pairs,
respectively. We categorize diagrams according to the $3$ parameters
$(k,\lambda,\sigma)$:
the maximum number of mutually crossing arcs, $k-1$,
the  minimum arc-length, $\lambda$ and the minimum stack-length,
$\sigma$. Here the length of an arc
$(i,j)$ is $j-i$ and a stack of length $\sigma$ is a sequence of
``parallel'' arcs of the form $((i,j),(i+1,j-1),\dots,(i+(\sigma-1),
j-(\sigma-1)))$, see Figure~\ref{F:lego1}. We call an arc of length
$\lambda$ a $\lambda$-arc.

%%%
%%%%%%%%%%%%%%%%%%%%%%%%%%%%%%%%%%%%%%%%%%%%%%%%%%%%%%%%%%%%%%%%%%%%%%%%%%
%%%
\begin{figure}[ht]
\centerline{%
\epsfig{file=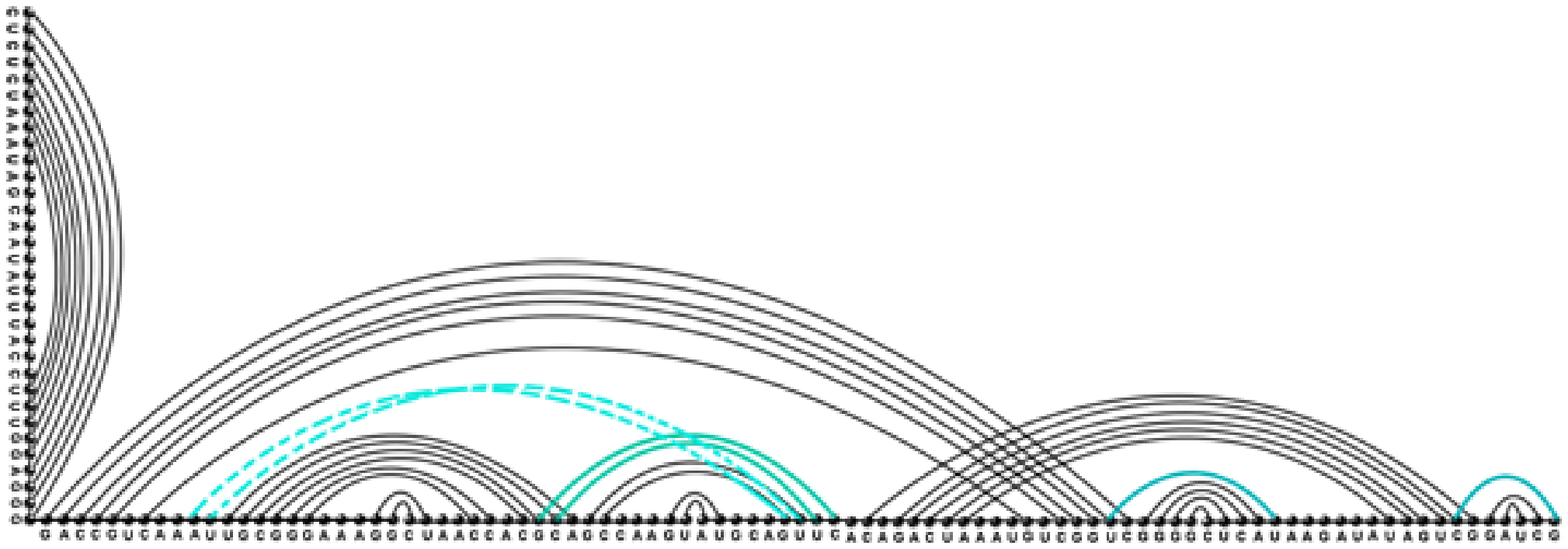,width=1.05\textwidth}\hskip15pt
 }
\caption{\small Diagram representation of the catalytic core region
of the group I self-splicing intron \cite{Chastain:92a}. Six
tertiary interactions shown in green lines. The gaps after
$\textbf{G}54, \textbf{U}72, \textbf{G}103$ and $\textbf{A}112$
indicate that some nucleotides are omitted which are involved in an
unrelated structural motif.
 }
\label{F:lego0}
\end{figure}

%%%
%%%%%%%%%%%%%%%%%%%%%%%%%%%%%%%%%%%%%%%%%%%%%%%%%%%%%%%%%%%%%%%%%%%%%%%%%%
%%%
In the following we call a $k$-noncrossing diagram with arc-length
$\ge 2$ and stack-length $\ge \sigma$ a $k$-noncrossing RNA
structure (of type $(k,\sigma)$). We denote the set (number) of
$k$-noncrossing RNA structures of type $(k,\sigma)$ by
$T_{k,\sigma}(n)$ (${\sf T}_{k,\sigma}^{}(n)$) and refer to
$k$-noncrossing  RNA structures for $k\ge 3$ as pseudoknot RNA
structures. Intuitively, a higher number of pairwise crossing arcs
is tantamount to higher structural complexity and crossing bonds are
reality \cite{Science:05a}. These pseudoknot bonds
\cite{Westhof:92a} occur in functional RNA (RNAseP
\cite{Loria:96a}), ribosomal RNA \cite{Konings:95a} and are
conserved in the catalytic core of group I introns, see
Figure~\ref{F:lego0}, where we show the diagram representation of
the catalytic core region of the group I self-splicing intron
\cite{Chastain:92a}. For $k=2$ we have RNA structures with no $2$
crossing arcs, i.e.~the well-known RNA secondary structures, whose
combinatorics was pioneered by Waterman {\it et.al.}
\cite{Penner:93c,Waterman:79a,Waterman:78a,Waterman:94a,Waterman:80}.
RNA secondary structures are structures of type $(2,1)$.

%%%
%%%%%%%%%%%%%%%%%%%%%%%%%%%%%%%%%%%%%%%%%%%%%%%%%%%%%%%%%%%%%%%%%%%%%%%%%%
%%%
%%%
%%%%%%%%%%%%%%%%%%%%%%%%%%%%%%%%%%%%%%%%%%%%%%%%%%%%%%%%%%%%%%%%%%%%%%%%%%
%%%

\begin{figure}[ht]
\centerline{%
\epsfig{file=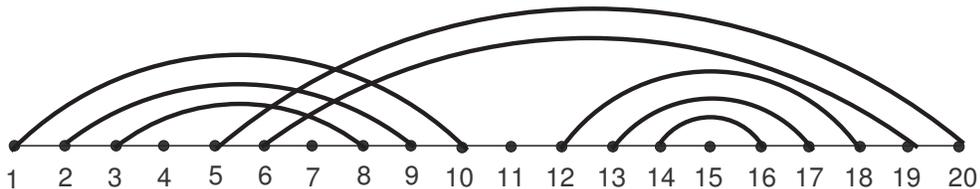,width=0.9\textwidth}\hskip15pt
 }
\caption{\small A canonical structure} \label{F:lego2a}
\end{figure}

There are many reasons why pseudoknot structures are fascinating.
First, compared to secondary structures their ``mathematical''
properties are much more intriguing
\cite{Reidys:07pseu,Reidys:07asy1,Reidys:07asy2}. Their enumeration
employs the nontrivial concepts of vacillating tableaux
\cite{Chen,Reidys:07vac} and singular expansions
\cite{Reidys:07asy1,Reidys:07asy2}. Secondly, the recurrence
relation for the numbers of $3$-noncrossing RNA \cite{Reidys:07pseu}
is, in contrast to that for secondary structures, ``enumerative''
but not  ``constructive''. This indicates that prediction of
pseudoknot RNA is much more involved compared to the dynamic
programming routine used for secondary structures.
Nevertheless, there exist several
prediction algorithms for RNA pseudoknot structures
\cite{Rivas:99a,Uemura:99a,Akutsu:00a,Lyngso:00a} which are able to
express certain ``types'' of pseudoknots. In this context the notion
of the ``language of RNA'' has been tossed \cite{Wang:07}. The
combinatorial analysis in
\cite{Reidys:07pseu,Reidys:07asy1,Reidys:07asy2} shows that
$3$-noncrossing RNA structures (${T}_{3,1}(n)$) exhibit an
exponential growth rate of $\frac{5+\sqrt{21}}{2}\approx 4.7913$ and
even when considering only structures with minimum arc-length $3$
the rate is $4.5492$. This
is bad news, since this rate exceeds already for $k=3$ the number of
sequences over the natural alphabet. Therefore, {\it a priori}, not
all $3$-noncrossing structures can be folded by sequences. The situation
becomes worse for higher $k$: the results of
\cite{Reidys:07asy1,Reidys:07asy2} imply the following exponential
growth rates for $k$-noncrossing RNA structures\footnote{here
$\gamma_{k,1}$ denotes the dominant real singularity of the generating function}
\begin{center}
\begin{tabular}{|c|c|c|c|c|c|c|c|c|c|}
%\hline
%  \multicolumn{9}{|c|}{\bf $k$-noncrossing RNA}\\
\hline $k$ & \small $2$ & \small $3$ & \small $4$ &\small $5$ &
\small $6$ & \small $7$ & \small $8 $
& \small $9$ & \small $10$\\
\hline $\gamma_{k,1}^{-1}$ & \small $2.6180$ & \small $4.7913$ &
\small$6.8540$ & \small $8.8873$  & \small $10.9087$ &
\small$12.9232$ & \small$14.9321$ & \small $16.9405$
& \small $18.9466$\\
\hline
\end{tabular}
\end{center}
Can we identify and analyze those $k$-noncrossing structures that {do}
``occur''?
To this end, let us consider this question in the
biophysical context: RNA structures are formed by Watson-Crick ({\bf
A-U}, {\bf G-C}) and ({\bf U-G}) base pairs and, due to the specific
chemistry of the latter, parallel bonds are thermodynamically more
stable. This fact is well-known and has led to the notion of
``canonical'' structures \cite{Stadler:07}, i.e.~structures in which
there exist {\it no} isolated base pair, see Figure~\ref{F:lego2a}.
%%%
%%%%%%%%%%%%%%%%%%%%%%%%%%%%%%%%%%%%%%%%%%%%%%%%%%%%%%%%%%%%%%%%%%%%%%%%%%
%%%
%%%
%%%
%%%%%%%%%%%%%%%%%%%%%%%%%%%%%%%%%%%%%%%%%%%%%%%%%%%%%%%%%%%%%%%%%%%%%%%%%%
%%%
The question then is, do canonical $k$-noncrossing structures
exhibit significantly smaller growth rates? Why this (to our
knowledge) has not been seriously pursued could be explained by a
result due to Schuster {\it et.al.} \cite{Dam:94}, who have proved
the following: the number of RNA secondary structures, ${\sf
T}_{2,1}(n)$, exhibits an exponential growth rate of
$\gamma_{2,1}^{-1}=2.6180$ while the number of canonical RNA
secondary structures ${\sf T}_{2,2}(n)$ has an exponential growth rate of
$\gamma_{2,2}^{-1}=1.9680$. In other words, the exponential growth
rate drops less than $25\%$ when passing from arbitrary to canonical
secondary structures. We remark that Schuster's enumerative result
is of central importance, since the growth rate of canonical
secondary structures
implies the existence of a ``many to one'' sequence to structure mapping.
This has, subsequently, led to the concept of neutral networks
\cite{Reidys:97}.

\begin{figure}[ht]
\centerline{%
\epsfig{file=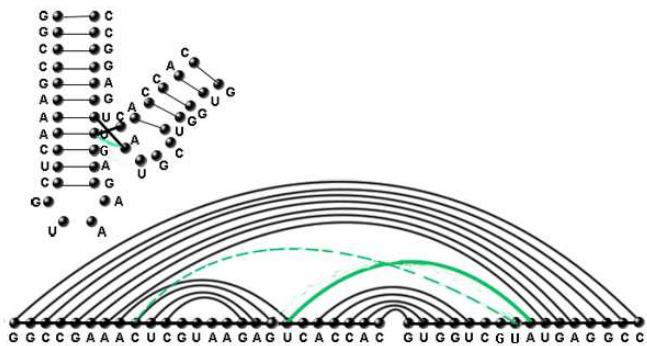,width=0.6\textwidth}\hskip15pt
 }
\caption{\small Biophysical constraints inducing parallel arcs: the
hammerhead ribozyme \cite{Batey}. Its two tertiary interactions are
shown in green lines, The gap after $\textbf{C}$25 indicates that
some nucleotides are omitted, which are involved in an unrelated
structural motif.
 }
\label{F:lego2}
\end{figure}

In the following we will develop a novel combinatorial framework which
allows to enumerate any RNA structure class of type $(k,\sigma)$, for any
$k,\sigma$. We then can report  good news: there is indeed a
significant drop in the exponential growth rates when passing from
$k$-noncrossing RNA structures to their canonical counterparts for
$k\ge 3$. Explicitly we can give the following data
\begin{center}
\begin{tabular}{|c|c|c|c|c|c|c|c|c|c|}
%\hline
%  \multicolumn{9}{|c|}{\textbf{$\lambda=2$}}\\
\hline $k$ & \small $2$ & \small $3$ & \small $4$ &\small $5$ &
\small $6$ & \small $7$ & \small $8 $
& \small$9$ & \small$10$\\
\hline $\gamma_{k,1}^{-1}$ & \small$2.6180$ & \small $4.7913$ &
\small$6.8540$ & \small $8.8873$  & \small $10.9087$ &
\small$12.9232$ & \small$14.9321$ & \small $16.9405$
& \small $18.9466$\\
\hline \small $\gamma_{k,2}^{-1}$ & \small $1.9680$ &\small
$2.5881$ & \small$3.0382$ & \small $3.4138$ & \small $3.7439$ &
\small$4.0420$ & \small $4.3159$
& \small$4.5714$ & \small$4.8114$\\
\hline
\end{tabular}
\end{center}
where the case $k=2$ is due to \cite{Dam:94}, which is independently confirmed
by our approach.
In particular, for $3$-noncrossing RNA structures, we have a drop in exponential
growth rate from $4.7913$ to $2.5881$, more than $46\%$ and for $k=10$ there is a
drop of more than $74\%$. As a result, the number of canonical $3$-noncrossing RNA
structures is very close to that of arbitrary secondary structures.
Intuitively this makes perfect sense since canonicity implies {\it parallel}
arcs which limits severely crossings and it can be expected to have dramatic
effect on $k$-noncrossing RNA for large $k$. In other words, the biophysical
constraints (thermodynamic stability) counteracts the combinatorial variety,
see Figure~\ref{F:lego2}

%%%
%%%%%%%%%%%%%%%%%%%%%%%%%%%%%%%%%%%%%%%%%%%%%%%%%%%%%%%%%%%%%%%%%%%%%%%%%%
%%%
%%
%%%
%%%%%%%%%%%%%%%%%%%%%%%%%%%%%%%%%%%%%%%%%%%%%%%%%%%%%%%%%%%%%%%%%%%%%%%%%%
%%%

\begin{figure}[ht]
\centerline{%
\epsfig{file=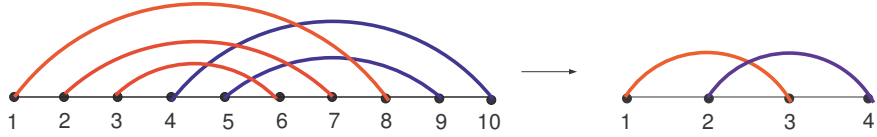,width=0.8\textwidth}\hskip15pt
 }
\caption{\small Core-structures. Each sequence of stacked arcs in the
$3$-noncrossing (canonical) structure (lhs) is replaced by its unique
arc with minimal
length (rhs). The so derived core-structure is unique. We show in
Lemma~\ref{L:core} that is assignment yields a well-defined mapping
(i.e.~no arcs of the form $(i,i+1)$ are being produced).}
\label{F:lego333}
\end{figure}

The main idea in this paper is to consider a new type of $k$-noncrossing
structure, that can be considered as being ``dual'' to canonical structures.
We consider $k$-noncrossing structures in which there
exists {\it no} two arcs of the form $(i,j),(i+1,j-1)$. These structures are
called $k$-noncrossing core-structures
and ${\sf C}_k^{}(n)$ denotes their number.
The key observation with respect to core-structures is the following:
any structure has a unique
core obtained by identifying all arcs contained in stacks by a single arc
and keeping isolated vertices.
Furthermore, the number of all structures is a sum of the numbers of the
corresponding core structures with {\it positive} integer coefficients. In
Figure~\ref{F:lego333} we illustrate the idea of how a core-structure is
obtained.
%%%
%%%%%%%%%%%%%%%%%%%%%%%%%%%%%%%%%%%%%%%%%%%%%%%%%%%%%%%%%%%%%%%%%%%%%%%%%%
%%%
%%
%%%
%%%%%%%%%%%%%%%%%%%%%%%%%%%%%%%%%%%%%%%%%%%%%%%%%%%%%%%%%%%%%%%%%%%%%%%%%%
%%%
It is of particular interest to note that Figure~\ref{F:lego333} shows that
deriving the core-structure can reduce the minimum arc-length, but cannot
produce arcs of the form $(i,i+1)$.
In Theorem~\ref{T:core} we derive the generating function
for core-structures which shows that ``most'' $k$-noncrossing structures
are in fact core-structures. Denoting the exponential growth rate of
$k$-noncrossing core-structures by $\kappa_k^{-1}$ we have the situation
\begin{center}
\begin{tabular}{|c|c|c|c|c|c|c|c|c|c|}
\hline $k$ & \small $2$ & \small $3$ & \small $4$ &\small $5$ &
\small $6$ & \small $7$ & \small $8 $
& \small$9$ & \small$10$\\
\hline $\gamma_{k,1}^{-1}$ & \small $2.6180$ & \small $4.7913$ &
\small$6.8540$ & \small $8.8873$  & \small $10.9087$ &
\small$12.9232$ & \small$14.9321$ & \small $16.9405$
& \small $18.9466$\\
\hline \small $\kappa_{k}^{-1}$ & \small$2.5152$ & \small $4.7097$
& \small$6.7921$ & \small$8.8378$  & \small $10.8672$ &
\small$12.8866$ & \small
$14.9031$& \small$16.9119$ & \small $18.9215$\\
\hline
\end{tabular}
\end{center}
In Theorem~\ref{T:arc-2} we derive a functional identity for the
generating function for $k$-noncrossing RNA structures with
stack-length $\ge \sigma$, which allows to obtain exact and
asymptotic results on ${\sf T}_{k,\sigma}^{}(n)$, i.e.~all
$k$-noncrossing RNA structures with stack-length $\ge \sigma$. In
its proof the number of $k$-noncrossing core-structures plays a
central role. As for the quality of the asymptotic expressions we
compare in the table below subexponential factors for $3$-and
$4$-noncrossing RNA structures with stack-length $\ge 2$,
$t_{3,2}(n)=\frac{311.2470\cdot 4!}{n(n-1)(n-2)(n-3)(n-4)}$
and $t_{4,2}(n)=\small 1.217\cdot 10^{7} n^{-\frac{21}{2}}$
with ${\sf T}_{3,2}(n)\,\gamma_{3,2}^{n}$ and ${\sf
T}_{4,2}(n)\,\gamma_{4,2}^{n}$, respectively. Here $\gamma_{k,\sigma}^{-1}$
denotes the respective exponential growth rate:
\begin{center}
\begin{tabular}{|c|c|c|c|c|}
\hline
  \multicolumn{5}{|c|}{\bf The subexponential factor}\\
\hline $n$ & ${\sf T}_{3,2}(n)\,\gamma_{3,2}^{n}$ &
$t_{3,2}(n)$ &
${\sf T}_{4,2}(n)\,\gamma_{4,2}^{n}$ & $t_{4,2}(n)$\\
\hline \small$50$ & \small $1.214\times 10^{-5}$ &
\small$2.938\times 10^{-5}$
& \small$3.115\times 10^{-8}$ & \small$1.763\times 10^{-7}$\\
\hline \small$60$ & \small $5.498\times 10^{-6}$ &
\small$1.140\times 10^{-5}$
& \small$6.884\times 10^{-9}$ & \small$2.599\times 10^{-8}$\\
\hline \small$70$ & \small $2.776\times 10^{-6}$ &
\small$5.143\times 10^{-6}$
& \small$1.841\times 10^{-9}$ & \small$5.151\times 10^{-9}$\\
\hline \small$80$ & \small $1.522\times 10^{-6}$ &
\small$2.589\times 10^{-6}$
& \small$5.708\times 10^{-10}$ & \small$1.268\times 10^{-9}$\\
\hline \small$90$ & \small $8.905\times 10^{-7}$ &
\small$1.416\times 10^{-6}$
& \small$1.991\times 10^{-10}$ & \small$3.680\times 10^{-9}$\\
\hline \small$100$ & \small $5.487\times 10^{-7}$ &
\small$8.268\times 10^{-7}$
& \small$7.650\times 10^{-11}$ & \small$1.217\times 10^{-10}$\\
\hline
\end{tabular}
\end{center}

%%%
%%%%%%%%%%%%%%%%%%%%%%%%%%%%%%%%%%%%%%%%%%%%%%%%%%%%%%%%%%%%%%%%%%%%%%%%%
%%%

\section{Some basic facts}\label{S:back}

%%%
%%%%%%%%%%%%%%%%%%%%%%%%%%%%%%%%%%%%%%%%%%%%%%%%%%%%%%%%%%%%%%%%%%%%%%%%%
%%%
In this Section we provide the basic facts needed for proving Theorem
\ref{T:core} in Section~\ref{S:core} and Theorem \ref{T:arc-2} in
Section~\ref{S:arc-2}. For background on crossings and nestings in
diagrams and partitions we recommend the paper of
Chen {\it et.al.}~\cite{Chen} and for the analytic combinatorics and
asymptotic analysis the book of Flajolet~\cite{Flajolet:07a}.
Our results are based on the generating function of $k$-noncrossing
RNA structures \cite{Reidys:07pseu}, and asymptotic analysis of
$k$-noncrossing RNA structures \cite{Reidys:07asy1,Reidys:07asy2},
summarized in Theorem~\ref{T:cool1} below.

\begin{figure}[ht]
\centerline{%
\epsfig{file=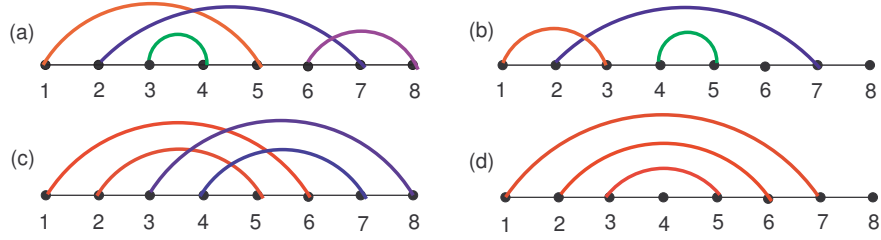,width=0.8\textwidth}\hskip15pt
 }
\caption{\small Basic diagram types:
(a) perfect matching ($f_3(8,0)$),
(b) partial matching with $1$-arc $(4,5)$ and isolated points $6,8$
($f_3(8,2)$),
(c) structure (i.e.~minimum arc-length $\ge 2$) with minimum stack-length $2$
and no isolated points (${\sf T}_{3,2}(8)$) and
(d) structure with minimum stack-length $3$ and isolated points $4,8$
(${\sf T}_{2,3}(8)$).}
\label{F:4-pic}
\end{figure}

Let us first recall our basic terminology, by ${T}_{k,\sigma}(n)$ we denote
the set of $k$-noncrossing RNA
structures with minimum stack length $\sigma$ and let ${\sf
T}_{k,\sigma}(n)$ denote their number. That is $T_{k,\sigma}(n)$ can
be identified with the set of diagrams with degree $\le 1$,
represented by drawing the vertices $1,\dots,n$ in a horizontal line
and its arcs $(i,j)$, where $i<j$, in the upper half plane with
arc-length $\ge 2$ and stack-length $\ge\sigma$, in which the
maximum number of mutually crossing arcs is $k-1$. Furthermore let
${T}_{k,\sigma}(n,h)$ denote the set of $k$-noncrossing RNA
structures stack-length $\ge \sigma$ having $h$ arcs and let ${\sf
T}_{k,\sigma}(n,h)$ denote their number.
We denote by $f_{k}(n,\ell)$ the number of $k$-noncrossing diagrams
with arbitrary arc-length and $\ell$ isolated points. In Figure~\ref{F:4-pic}
we display the various types of diagrams involved.
%%%
%%%%%%%%%%%%%%%%%%%%%%%%%%%%%%%%%%%%%%%%%%%%%%%%%%%%%%%%%%%%%%%%%%%%%%%%%%
%%%
%%
%%%
%%%%%%%%%%%%%%%%%%%%%%%%%%%%%%%%%%%%%%%%%%%%%%%%%%%%%%%%%%%%%%%%%%%%%%%%%%
%%%

The following identities are due to Grabiner {\it et. al.}
\cite{Grabiner:93a}
\begin{eqnarray}\label{E:ww0}
\label{E:ww1}
\sum_{n\ge 0} f_{k}(n,0)\cdot\frac{x^{n}}{n!} & = &
\det[I_{i-j}(2x)-I_{i+j}(2x)]|_{i,j=1}^{k-1} \\
\label{E:ww2}
\sum_{n\ge 0}\left\{\sum_{\ell=0}^nf_{k}(n,\ell)\right\}\cdot\frac{x^{n}}{n!}
&= &
e^{x}\det[I_{i-j}(2x)-I_{i+j}(2x)]|_{i,j=1}^{k-1}
\end{eqnarray}
where $I_{r}(2x)=\sum_{j \ge 0}\frac{x^{2j+r}}{{j!(r+j)!}}$ denotes
the hyperbolic Bessel function of the first kind of order $r$.
Eq.~(\ref{E:ww1})
and (\ref{E:ww2}) allow ``in principle'' for explicit computation of
the numbers $f_k(n,\ell)$. In particular for $k=2$ and $k=3$ we have
the formulas
\begin{equation}\label{E:2-3}
f_2(n,\ell)  =  \binom{n}{\ell}\,C_{(n-\ell)/2}\quad
\text{\rm and}\quad  f_{3}(n,\ell)=
{n \choose \ell}\left[C_{\frac{n-\ell}{2}+2}C_{\frac{n-\ell}{2}}-
      C_{\frac{n-\ell}{2}+1}^{2}\right] \ ,
\end{equation}
where $C_m$ denotes the $m$-th Catalan number. The second formula results from
a determinant formula enumerating pairs of nonintersecting Dyck-paths.
In view of
$$
f_{k}(n,\ell) ={n \choose \ell} f_{k}(n-\ell,0)
$$
everything can be reduced to perfect matchings, where we have the
following situation: there exists an asymptotic approximation of the
hyperbolic Bessel function due to \cite{Odlyzko:95a} and employing
the subtraction of singularities-principle \cite{Odlyzko:95a} one
can prove
\begin{equation}\label{E:f-k-imp}
\forall\, k\in\mathbb{N};\qquad  f_k(2n,0)\sim \varphi_k(n)
\left(\frac{1}{\rho_k}\right)^n\  ,
\end{equation}
where $\rho_k$ is the dominant real singularity of $\sum_{n\ge
0}f_k(2n,0)z^n$ and $\varphi_k(n)$ is a polynomial over $n$.
Via Hadamard's formula, $\rho_k$ can be expressed as
\begin{equation}\label{E:rho-k}
\rho_k=\lim_{n\to \infty}(f_k(2n,0))^{-\frac{1}{2n}} \ .
\end{equation}
Eq.~(\ref{E:f-k-imp}) allows for any $k$ to obtain $\varphi_k(n)$, explicitly.

As for the generating function and asymptotics of $k$-noncrossing RNA
structures
we have the following result
%%%
%%%%%%%%%%%%%%%%%%%%%%%%%%%%%%%%%%%%%%%%%%%%%%%%%%%%%%%%%%%%%%%%%%%%%%%%%
%%%
\begin{theorem}\label{T:cool1}\cite{Reidys:07pseu,Reidys:07asy1}
Let $k\in\mathbb{N}$, $k\ge 2$. Then the number of $k$-noncrossing RNA
structures with $\frac{n-\ell}{2})$ arcs,
${\sf T}_{k,1}(n,\frac{n-\ell}{2})$ and the number of $k$-noncrossing RNA
structures, ${\sf T}_{k,1}(n)$ are given by
\begin{eqnarray}\label{E:ddd}
{\sf T}_{k,1}(n,\frac{n-\ell}{2})
& = & \sum_{b=0}^{\lfloor n/2\rfloor}(-1)^{b}{n-b \choose b}
f_{k}(n-2b,\ell) \\
\label{E:sum}
{\sf T}_{k,1}(n)
& = & \sum_{b=0}^{\lfloor n/2\rfloor}(-1)^{b}{n-b \choose b}
\left\{\sum_{\ell=0}^{n-2b}f_{k}(n-2b,\ell)\right\} \ ,
\end{eqnarray}
where $\{\sum_{\ell=0}^{n-2b}f_{k}(n-2b,\ell)\}$ is given via
eq.~{\rm (\ref{E:ww2})} and furthermore
\begin{eqnarray*}
\label{E:konk3} {\sf T}_{3,1}(n) & \sim & \frac{10.4724\cdot
4!}{n(n-1)\dots(n-4)}\,
\left(\frac{5+\sqrt{21}}{2}\right)^n \ .\\
\end{eqnarray*}
\end{theorem}
%%%
%%%%%%%%%%%%%%%%%%%%%%%%%%%%%%%%%%%%%%%%%%%%%%%%%%%%%%%%%%%%%%%%%%%%%%%%%
%%%

The following functional identity is due to \cite{Reidys:07asy1} and relates
the bivariate generating function for ${\sf T}_{k,1}(n,h)$, the number of
RNA pseudoknot structures with $h$ arcs to the generating
function of $k$-noncrossing perfect matchings.

%%%
%%%%%%%%%%%%%%%%%%%%%%%%%%%%%%%%%%%%%%%%%%%%%%%%%%%%%%%%%%%%%%%%%%%%%%%%%
%%%
\begin{lemma}\label{L:arc-2}
Let $k\in\mathbb{N}$, $k\ge 2$ and $z,u$ be indeterminants over
$\mathbb{C}$. Then we have
\begin{equation}\label{E:rr}
\sum_{n\ge 0} \sum_{h\le n/2} {\sf T}_{k,1}(n,h) \ u^{2h} z^n =
\frac{1}{u^2z^2-z+1}
\sum_{n\ge 0} f_k(2n,0) \left(\frac{uz}{u^2z^2-z+1}\right)^{2n} \ .
\end{equation}
In particular we have for $u=1$,
\begin{equation}\label{E:oha}
\sum_{n\ge 0} {\sf T}_{k,1}(n) \ z^{n} =
\frac{1}{z^2-z+1}\,
\sum_{n\ge 0} f_k(2n,0) \left(\frac{z}{z^2-z+1}\right)^{2n} \ .
\end{equation}
\end{lemma}
%%%
%%%%%%%%%%%%%%%%%%%%%%%%%%%%%%%%%%%%%%%%%%%%%%%%%%%%%%%%%%%%%%%%%%%%%%%%%%%
%%%

In view of Lemma~\ref{L:arc-2} it is of interest to deduce
relations between the coefficients from the equality of generating
functions. The class of theorems that deal with this deduction are
called transfer-theorems \cite{Flajolet:07a}. One key ingredient in
this framework is a specific domain in which the functions in
question are analytic, which is ``slightly'' bigger than their
respective radius of convergence. It is tailored for extracting the
coefficients via Cauchy's integral formula. Details on the method
can be found in \cite{Flajolet:07a} and its application to
$3$-noncrossing RNA in \cite{Reidys:07asy1}. To be precise the domain
in question is
%%%
%%%%%%%%%%%%%%%%%%%%%%%%%%%%%%%%%%%%%%%%%%%%%%%%%%%%%%%%%%%%%%%%%%%%%%%%%
%%%
\begin{definition}\label{D:delta}
Given two numbers $\phi,R$, where $R>1$ and $0<\phi<\frac{\pi}{2}$ and
$\rho\in\mathbb{R}$ the open domain $\Delta_\rho(\phi,R)$ is defined as
\begin{equation}
\Delta_\rho(\phi,R)=\{ z\mid \vert z\vert < R, z\neq \rho,\,
\vert {\rm Arg}(z-\rho)\vert >\phi\}
\end{equation}
A domain is a $\Delta_\rho$-domain if it is of the form
$\Delta_\rho(\phi,R)$ for some $R$ and $\phi$.
A function is $\Delta_\rho$-analytic if it is analytic in some
$\Delta_\rho$-domain.
\end{definition}
%%%
%%%%%%%%%%%%%%%%%%%%%%%%%%%%%%%%%%%%%%%%%%%%%%%%%%%%%%%%%%%%%%%%%%%%%%%%%
%%%
%%%%
%%%%%%%%%%%%%%%%%%%%%%%%%%%%%%%%%%%%%%%%%%%%%%%%%%%%%%%%%%%%%%%%%%%%%%%%%%
%%%%

 We use the notation
\begin{equation}\label{E:genau}
\left(f(z)=O\left(g(z)\right) \
\text{\rm as $z\rightarrow \rho$}\right)\quad \Longleftrightarrow \quad
\left(f(z)/g(z) \ \text{\rm is bounded as $z\rightarrow \rho$}\right)
\end{equation}
and if we write $f(z)=O(g(z))$ it is implicitly assumed that $z$
tends to a (unique) singularity. $[z^n]\,f(z)$ denotes the
coefficient of $z^n$ in the power series expansion of $f(z)$ around
$0$.

%%%
%%%%%%%%%%%%%%%%%%%%%%%%%%%%%%%%%%%%%%%%%%%%%%%%%%%%%%%%%%%%%%%%%%%%%%%%%
%%%
\begin{theorem}\label{T:transfer}\cite{Flajolet:05}
Let $f(z),g(z)$ be a $\Delta_\rho$-analytic functions given by power
series $f(z)=\sum_{n\ge 0} a_nz^n$ and $g(z)=\sum_{n\ge 0}b_n z^n$.
Suppose $f(z)= O(g(z))$ for all $z\in\Delta_\rho$ and $b_n\sim
\varphi(n)(\rho^{-1})^n$, where $\varphi(n)$ is a polynomial over
$n$. Then
\begin{equation}
a_n = [z^n]\, f(z)  \sim  K \ [z^n]\, g(z)= K\,b_n\sim K\,
\varphi(n)(\rho^{-1})^n
\end{equation}
for some constant $K$.
\end{theorem}

Transfer theorems are accordingly a translation of error terms from
functions to coefficients and guaranteed when the functions in question
are analytic in some $\Delta_\rho$-domain.
%%%
%%%%%%%%%%%%%%%%%%%%%%%%%%%%%%%%%%%%%%%%%%%%%%%%%%%%%%%%%%%%%%%%%%%%%%%%%
%%%

%%%
%%%%%%%%%%%%%%%%%%%%%%%%%%%%%%%%%%%%%%%%%%%%%%%%%%%%%%%%%%%%%%%%%%%%%%%%%
%%%

\section{Core-structures}\label{S:core}

%%%
%%%%%%%%%%%%%%%%%%%%%%%%%%%%%%%%%%%%%%%%%%%%%%%%%%%%%%%%%%%%%%%%%%%%%%%%%
%%%

As discussed in the Introduction, a {core-structure} is a
$k$-noncrossing structure with no stacked base pairs. We denote the
set and number of core-structures over $[n]$ by ${C}_{k}(n)$ and
${\sf C}_{k}(n)$, respectively. Analogously ${C}_{k}(n,h)$ and ${\sf
C}_{k}(n,h)$ denote the set and the number of core-structures having
$h$ arcs.
In Lemma~\ref{L:core} below we establish that the
number of all $k$-noncrossing structures with stack-length $\ge
\sigma$ is a sum of the numbers of $k$-noncrossing cores with
positive integer coefficients.

\begin{figure}[ht]
\centerline{%
\epsfig{file=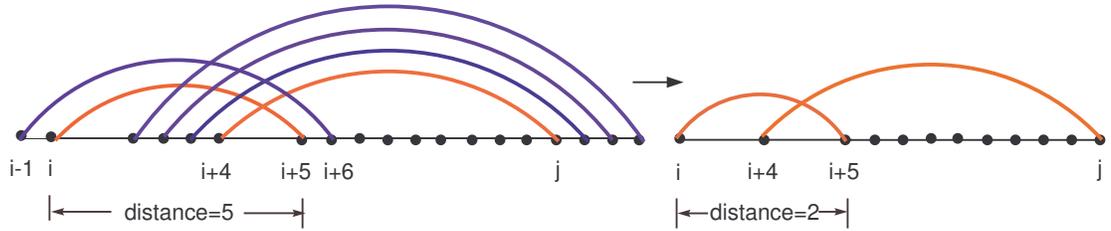,width=1\textwidth}\hskip15pt
 }
\caption{\small Core-structures will in general have $2$-arcs:
the structure $\delta\in T_{3,2}(19)$ (lhs) is mapped into its core
$c(\delta)$ (rhs).
Clearly $\delta$ has arc-length $\ge 5$ and as a consequence of the
collapse of the stack $((i+1,j+3),\dots,(i+4,j))$ (the blue arcs are being
removed) into the arc $(i+4,j)$, $c(\delta)$ contains the $2$-arc $(i,i+5)$.
}
 \label{F:lego4}
\end{figure}

%%%
%%%%%%%%%%%%%%%%%%%%%%%%%%%%%%%%%%%%%%%%%%%%%%%%%%%%%%%%%%%%%%%%%%%%%%
%%%
\begin{lemma}\label{L:core}{\bf (Core-lemma)}
For $k,h,\sigma\in \mathbb{N}$, $k\ge 2$, $1\le h\le n/2$ we have
\begin{equation}\label{E:relation1}
{\sf T}_{k,\sigma}^{}(n,h)=\sum_{b=\sigma-1}^{h-1}
{b+(2-\sigma)(h-b)-1 \choose h-b-1}{\sf C}_{k}^{}(n-2b,h-b) \ .
\end{equation}
\end{lemma}
%%%
%%%%%%%%%%%%%%%%%%%%%%%%%%%%%%%%%%%%%%%%%%%%%%%%%%%%%%%%%%%%%%%%%%%%%%
%%%
\begin{remark}\label{R:0}{\rm
Lemma~\ref{L:core} cannot be used in order to enumerate
diagrams with arc-length $\ge \lambda$, where $\lambda>2$ and stack-length
$\sigma$. Basically, $k$-noncrossing structures
with arc-length $\ge \lambda$ have core-structures with
arc-length $2$, see Figure~\ref{F:lego4}. The enumeration of
$k$-noncrossing RNA structures with arc-length $\ge 3$ and stack-length
$\ge 2$ is work in progress.}
\end{remark}
%%%
%%%%%%%%%%%%%%%%%%%%%%%%%%%%%%%%%%%%%%%%%%%%%%%%%%%%%%%%%%%%%%%%%%%%%%%%%%%
%%%
%%%
%%%%%%%%%%%%%%%%%%%%%%%%%%%%%%%%%%%%%%%%%%%%%%%%%%%%%%%%%%%%%%%%%%%%%%%%%%
%%%
%%%
%%%%%%%%%%%%%%%%%%%%%%%%%%%%%%%%%%%%%%%%%%%%%%%%%%%%%%%%%%%%%%%%%%%%%%%%%%
%%%
\begin{proof}
First, there exists a mapping from $k$-noncrossing structures with $h$ arcs
and minimum stack size $\sigma$ over $[n]$ into core-structures:
\begin{equation}
c\colon T_{k,\sigma}(n,h)\rightarrow
\dot\bigcup_{0\le b\le h-1}{ C}_k(n-2b,h-b), \quad
\delta\mapsto c(\delta)
\end{equation}
where the core-structure $c(\delta)$ is obtained in two steps: first we
map arcs and isolated vertices as follows
\begin{equation}
\forall \,\ell\ge \sigma-1;\quad
((i-\ell,j+\ell),\dots,(i,j)) \mapsto (i,j) \ \quad \text{\rm and} \quad
j \mapsto j \quad \text{\rm if $j$ is isolated.}
\end{equation}
and second we relabel the vertices of the resulting diagram from left to
right in increasing order. That is we replace each stack by a single arc
and keep isolated points and then relabel, see Figure~\ref{F:lego5}.
%%%
%%%%%%%%%%%%%%%%%%%%%%%%%%%%%%%%%%%%%%%%%%%%%%%%%%%%%%%%%%%%%%%%%%%%%%%%%%
%%%
\begin{figure}[ht]
\centerline{%
\epsfig{file=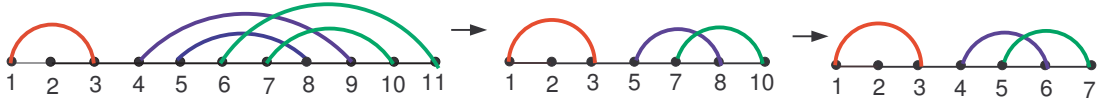,width=1\textwidth}\hskip15pt
 }
\caption{\small The mapping $c\colon T_{k,\sigma}(n,h)\longrightarrow
\dot\bigcup_{0\le b\le h-1}{C}_k(n-2b,h-b)$ is obtained in two steps:
first contraction of the stacks and secondly relabeling of the resulting
diagram.
}
 \label{F:lego5}
\end{figure}
%%%
%%%%%%%%%%%%%%%%%%%%%%%%%%%%%%%%%%%%%%%%%%%%%%%%%%%%%%%%%%%%%%%%%%%%%%%%%%
%%%
We have to prove that $c\colon T_{k,\sigma}(n,h)\longrightarrow
\dot\bigcup_{0\le b\le h-1}{C}_k(n-2b,h-b)$ is well-defined,
i.e.~that $c$ cannot produce
$1$-arcs. Indeed, since $\delta\in T_{k,\sigma}(n,h)$, $\delta$ does not
contain $1$-arcs we can conclude that $c(\delta)$ has by construction arcs
of length $\ge 2$. $c$ is by construction surjective.
Keeping track of multiplicities gives rise to the
map
\begin{equation}\label{E:core}
f_{k,\sigma}^{} \colon T^{}_{k,\sigma}(n,h) \rightarrow
\dot\bigcup_{0\le b\le h-1} \left[C_k^{}(n-2b,h-b) \times
\left\{(a_{j})_{1\le j\le h-b}\mid\sum_{j=1}^{h-b}a_{j}=b, \
a_{j}\ge \sigma-1 \right\}\right],
\end{equation}
given by $f_{k,\sigma}^{ }(\delta)=(c(\delta),(a_{j})_{1\le j\le
h-b})$.
We can conclude that $f_{k,\sigma}^{ }$ is well-defined and a bijection.
We proceed computing the multiplicities of the resulting core-structures:\\
{\it Claim.}
\begin{equation}\label{E:core-1}
\vert\{(a_j)_{1\le j\le h-b}\mid \sum_{j=1}^{h-b}a_j=b; \ a_j\ge
\sigma -1\}\vert = {b+(2-\sigma)(h-b)-1 \choose h-b-1} \ .
\end{equation}
Clearly, $a_j\ge \sigma -1$ is equivalent to $\mu_{j}=a_{j}-\sigma+2\ge 1$
and we have
$$
\sum_{j=1}^{h-b}\mu_{j}=\sum_{j=1}^{h-b}(a_{j}-\sigma+2)=b+(2-\sigma)(h-b)
\ .
$$
We next show that
\begin{equation}
\vert\{(\mu_{j})_{1\le j \le h-b}\}|\sum_{j=1}^{h-b}\mu_{j}=
b+(2-\sigma)(h-b); \, \mu_{j}\ge 1\vert
\end{equation}
is equal the number of $(h-b-1)$-subset in $\{1,2,\ldots,b+
(2-\sigma)(h-b)-1\}$. Consider the set
\begin{equation}
\{\mu_{1},\mu_{1}+\mu_{2},\ldots,\mu_{1}+\mu_{2}+\cdots+\mu_{h-b-1}\}
\end{equation}
consisting of $h-b-1$
distinct elements of $[b+(2-\sigma)(h-b)-1]=\{1,2,\ldots ,b+(2-\sigma)
(h-b)-1\}$. Therefore $\{\mu_{1},\mu_{1}+\mu_{2},\ldots,\mu_{1}+
\mu_{2}+\cdots+\mu_{h-b-1}\}$ is a $(h-b-1)$-subset of
$[b+(2-\sigma)(h-b)-1]$. Given any $(h-b-1)$-subset of
$[b+(2-\sigma)(h-b)-1]$, we can arrange its elements in linear order
and retrieve the sequence $\{\mu_{i}|\; 1\le i\le h-b\}$ of positive
integers with sum $b+(2-\sigma)(h-b)$. Therefore the above assignment is
a bijection. Since the number of $(h-b-1)$-subsets of $[b+(2-\sigma)
(h-b)-1]$ is given by ${b+(2-\sigma)(h-b)-1 \choose h-b-1}$ the Claim
follows.\\
We can conclude from the Claim and eq.~(\ref{E:core}) that
\begin{equation}
{\sf T}^{ }_{k,\sigma}(n,h)=\sum_{b=\sigma-1}^{h-1}
{b+(2-\sigma)(h-b)-1 \choose h-b-1}{\sf C}^{ }_{k}(n-2b,h-b)
\end{equation}
holds and the lemma follows.
\end{proof}

Next, we prove a functional identity between the bivariate generating
functions of ${\sf T}^{ }_{k,\sigma}(n,h)$ and ${\sf C}^{}_{k}(n,h)$.
This identity plays a central role in proving Theorem~\ref{T:core} and
Theorem~\ref{T:arc-2} in Section~\ref{S:arc-2}.

%%%
%%%%%%%%%%%%%%%%%%%%%%%%%%%%%%%%%%%%%%%%%%%%%%%%%%%%%%%%%%%%%%%%%%%%%%
%%%
\begin{lemma}{\bf }\label{L:www}
Let $k,\sigma\in \mathbb{N}$, $k\ge 2$ and let $u,x$ be indeterminants.
Then we have the functional relation
\begin{equation}\label{E:f1a}
\sum_{n \ge 0}\sum_{h\le \frac{n}{2}}{\sf T}^{ }_{k,
\sigma}(n,h)u^hx^n =
\sum_{n \ge 0}\sum_{h \le \frac{n}{2}}{\sf
C}^{ }_{k}(n,h)\left(\frac{u\cdot (ux^2)^{
\sigma-1}}{1-ux^2}\right)^hx^n+\frac{x}{1-x}
\end{equation}
and in particular, for $u=1$
\begin{equation}\label{E:f1b}
\sum_{n \ge 0} {\sf T}^{ }_{k,\sigma}(n)x^n =
\sum_{n \ge 0}\sum_{h \le \frac{n}{2}}{\sf
C}^{ }_{k}(n,h)\left(\frac{ (x^2)^{\sigma-1}}{1-x^2}\right)^hx^n  +
\frac{x}{1-x}\ .
\end{equation}
\end{lemma}
%%%
%%%%%%%%%%%%%%%%%%%%%%%%%%%%%%%%%%%%%%%%%%%%%%%%%%%%%%%%%%%%%%%%%%%%%%
%%%

\begin{proof}
We set $\sum_{n \ge 0}\sum_{h\le\frac{n}{2}}{\sf C}_{k}^{ }
(n,h)u^hx^n=\sum_{h\ge 0}\varphi_{h}^{ }(x)u^h$ and compute
\begin{equation}
\sum_{n \ge 0}\sum_{h\le \frac{n}{2}}
{\sf T}_{k,\sigma}^{ }(n,h)u^hx^n
 =
\sum_{n \ge 0}\sum_{h \le \frac{n}{2}}\sum_{b\le h-1}{\sf
C}_{k}^{ }(n-2b,h-b){b+(2-\sigma)(h-b)-1 \choose
h-b-1}u^h x^n + \sum_{i\ge 1}x^i
\end{equation}
where the term $\sum_{i\ge 1}x^i=\frac{x}{1-x}$ comes from the fact that
for $h=0$ the binomial
$$
{b+(2-\sigma)(h-b)-1 \choose h-b-1}
$$
is zero, while the lhs counts for any $i\ge 1$ the unique structure
having only isolated vertices. We proceed to compute
\begin{eqnarray*}
&= & \sum_{h \ge 0}\sum_{b\le h-1}\sum_{n \ge 2h}{\sf
C}_{k}^{ }(n-2b,h-b)x^{n-2b}{b+(2-\sigma)(h-b)-1
\choose h-b-1}u^h x^{2b}+ \frac{x}{1-x}\\
&=& \sum_{b \ge 0}\sum_{b\le
h}\varphi_{h-b}^{ }(x){b+(2-\sigma)(h-b)-1 \choose h-b-1}u^h
x^{2b} + \frac{x}{1-x} \ .
\end{eqnarray*}
Setting $m=h-b$ and subsequently interchanging the summation indices we
arrive at
\begin{eqnarray*}
\sum_{n \ge 0}\sum_{h\le \frac{n}{2}}
{\sf T}_{k,\sigma}^{ }(n,h)u^hx^n
&= & \sum_{b \ge 0}\sum_{1\le m}\varphi_{m}^{ }(x){
b+(2-\sigma)m-1\choose m-1}u^m(ux^2)^b + \frac{x}{1-x}\\
&=& \sum_{m \ge
0}\varphi_{m}^{ }(x)\left(\frac{u\cdot(ux^2)^{\sigma-1}}
{1-ux^2}\right)^m + \frac{x}{1-x}\\
&=& \sum_{n \ge 0}\sum_{h \le \frac{n}{2}}{\sf
C}_{k}^{ }(n,h)\left(\frac{u\cdot
(ux^2)^{\sigma-1}}{1-ux^2}\right)^hx^n + \frac{x}{1-x} \ ,
\end{eqnarray*}
whence Lemma~\ref{L:www}.
\end{proof}
We next enumerate core-structures. The Theorem has two main parts, the first
is the ``inversion'' of Lemma~\ref{L:core}. It allows to express
core-structures
via all structures and follows by M\"obius inversion. The second part deals
with the asymptotics of core-structures.
The asymptotic formula follows then from transfer Theorems (the super-critical
case) \cite{Flajolet:07a} applied to some version of the functional
identity of Lemma~\ref{L:arc-2}.

%%%
%%%%%%%%%%%%%%%%%%%%%%%%%%%%%%%%%%%%%%%%%%%%%%%%%%%%%%%%%%%%%%%%%%%%%%
%%%
\begin{theorem}{\bf (Core-structures)}\label{T:core}
Suppose $k\in\mathbb{N}$, $k\ge 2$, let $x$ be an indeterminant,
$\rho_k$ the dominant, positive real singularity of
$\sum_{n\ge 0}f_k(2n,0)z^n$ (eq.~(\ref{E:rho-k})) and
$u_1(x)=\frac{1}{1+x^2}$.
Then for $h\ge 1$, the numbers of $k$-noncrossing
core-structures, ${\sf C}_k(n)$ are given by
\begin{equation}\label{E:inversion}
{\sf C}_{k}^{ }(n,h)=
\sum_{b=0}^{h-1}(-1)^{h-b-1}{h-1 \choose b}
{\sf T}^{ }_{k,1}(n-2h+2b+2,b+1) \ .
\end{equation}
Furthermore we have the functional equation
\begin{eqnarray}\label{E:C1}
\sum_{n \ge 0}{\sf C}_{k}^{}(n)\ x^n & = &
\frac{1}{u_1x^2-x+1}\sum_{n \ge
0}f_{k}(2n,0)\left(\frac{\sqrt{u_1}x}{u_1x^2-x+1}\right)^{2n}
- \frac{x}{1-x}
\end{eqnarray}
and the asymptotic expression
\begin{equation}\label{E:growth-free}
{\sf C}_{k}^{}(n)\sim \varphi_k(n)\ \left(
\frac{1}{\kappa_{k}}\right)^n \ ,
\end{equation}
where $\kappa_{k}$ is the dominant positive real singularity of
$\sum_{n \ge 0}{\sf C}_{k}^{}(n)x^n$ and the minimal positive
real solution of the equation
$\frac{\sqrt{u_1}\ x}{u_1x^2-x+1}  =  \rho_k$ and $\varphi_k(n)$ is
a polynomial over $n$ derived from the asymptotic expression of
$f_k(2n,0)\sim \varphi_k(n)\left(\frac{1} {\rho_k}\right)^n$ of
eq.~(\ref{E:f-k-imp}).
\end{theorem}
%%%
%%%%%%%%%%%%%%%%%%%%%%%%%%%%%%%%%%%%%%%%%%%%%%%%%%%%%%%%%%%%%%%%%%%%%%
%%%
\begin{center}
\begin{tabular}{c|ccccccccccccccc}
%\hline
%  \multicolumn{9}{|c|}{\bf $k$-noncrossing RNA}\\
$n$ & \small $1$ & \small $2$ & \small $3$ & \small $4$ & \small $5$
& \small $6$ & \small $7$ & \small$8$ & \small$9$ & \small$10$ &
\small $11$ & \small $12$ & \small $13$ &\small $14$ & \small $15$\\
\hline${\sf C}_{3}(n)$ & \small $1$ & \small $1$ & \small$2$ &
\small $5$  & \small $12$ & \small$31$ & \small$88$ & \small $263$ &
\small $814$ & \small $2604$ & \small $8575$ & \small $28936$
&\small $99726$ &\small $350151$ & \small$1249865$\\
${\sf C}_{4}(n)$ & \small $1$ & \small $1$ & \small$2$ & \small $5$
& \small $12$ & \small$32$ & \small$95$ & \small $301$ & \small
$1001$ & \small $3495$ & \small $12708$ & \small $47932$
& \small $186581$ & \small $747619$ &\small $3073207$\\
\end{tabular}
\end{center}
\begin{proof}
We set
\begin{eqnarray*}
\forall\, 0\le i\le h-1;\qquad  a(i) & = &
{\sf C}_k^{}(n-2(h-1-i),i+1) \\
\forall\, 0\le i\le h-1;\qquad  b(i) & = & {\sf T}_{k,1}^{ }
(n-2(h-1-i),i+1) \ .
\end{eqnarray*}
We first employ Lemma~\ref{L:core} for $\sigma=1$:
$$
{\sf T}^{ }_{k}(n,h)=\sum_{b=0}^{h-1}{h-1 \choose b}\, {\sf C}^{
}_{k}(n-2b,h-b) \quad \Longleftrightarrow \quad
b(h-1)=\sum_{i=0}^{h-1}\binom{h-1}{i}\, a(i) .
$$
Via M\"obius-inversion we arrive at $a(h-1)=\sum_{i=0}^{h-1}(-1)^{h-1-i}\,
\binom{h-1}{i}b(i)$, which is equivalent to
\begin{equation}\label{E:Moebius}
{\sf C}_{k}^{ }(n,h) = \sum_{b=0}^{h-1}(-1)^{h-b-1}{h-1 \choose b}
{\sf T}^{ }_{k,1}(n-2h+2b+2,b+1) \ ,
\end{equation}
whence eq.~(\ref{E:inversion}).
We proceed by proving eq.~(\ref{E:C1}). First Lemma~\ref{L:www} implies:
\begin{eqnarray}\label{E:kk}
\sum_{n \ge 0}\sum_{h\le \frac{n}{2}}{\sf T}_{k,1}^{ }(n,h)u^hx^n & = &
\sum_{n \ge 0}\sum_{h \le \frac{n}{2}}{\sf C}_{k}^{ }(n,h)
\left(\frac{u}{1-ux^2}\right)^hx^n + \frac{x}{1-x}
\end{eqnarray}
and we inspect that $u_1(x)=\frac{1}{1+x^2}$ is the unique solution for
$\frac{u}{1-ux^2}=1$. Accordingly we obtain
\begin{eqnarray*}
\sum_{n \ge 0}\sum_{h\le \frac{n}{2}}{\sf T}_{k,1}^{ }
(n,h)u_1^h\,x^n & = &
\sum_{n \ge 0}{\sf C}_{k}^{ }(n)\, x^n + \frac{x}{1-x}\ .
\end{eqnarray*}
Secondly, setting $u=\sqrt{u_1}$, Lemma~\ref{L:arc-2} provides an
interpretation of the lhs of eq.~(\ref{E:kk}):
\begin{equation}
\sum_{n\ge 0} \sum_{h\le n/2} {\sf T}_{k,1}^{}(n,h) \ u_1^{h} x^n =
\frac{1}{u_1x^2-x+1}
\sum_{n\ge 0} f_k(2n,0) \left(\frac{\sqrt{u_1}x}{u_1x^2-x+1}\right)^{2n}
\end{equation}
and we can conclude
\begin{eqnarray*}
\sum_{n \ge 0}{\sf C}_{k}^{}(n)\, x^n & = &
\sum_{n \ge 0}\sum_{h\le \frac{n}{2}}{\sf T}_{k,1}^{}(n,h)u_1^hx^n
- \frac{x}{1-x}\\
& = & \frac{1}{u_1x^2-x+1}\sum_{n \ge
0}f_{k}(2n,0)\left(\frac{\sqrt{u_1}x}{u_1x^2-x+1}\right)^{2n}
-\frac{x}{1-x}\ ,
\end{eqnarray*}
whence eq~(\ref{E:C1}).
As for eq.~(\ref{E:growth-free}) we consider the functional equation
$$
\sum_{n \ge 0}{\sf C}_{k}^{}(n)x^n =
\frac{1}{u_1x^2-x+1}\sum_{n \ge
0}f_{k}(2n,0)\left(\frac{\sqrt{u_1}x}{u_1x^2-x+1}\right)^{2n}
-\frac{x}{1-x}\ .
$$
Let us denote $W(x)=\sum_{n \ge 0}f_{k}(2n,0)\left(\frac{\sqrt{u_1}x}
             {u_1x^2-x+1}\right)^{2n}$.\\
{\it Claim.} All dominant singularities of $\sum_{n \ge 0}{\sf C}_{k}
             (n)\, z^n$ are dominant singularities of $W(z)$
             and $\kappa_k$ is a dominant singularity. \\
To prove the Claim we observe that a dominant singularity of
$$
\frac{1}{u_1z^2-z+1}\sum_{n \ge 0}f_{k}(2n,0)
\left(\frac{\sqrt{u_1}z}{u_1z^2-z+1}\right)^{2n}
- \frac{z}{1-z}
$$
is either a singularity of $W(z)$ or $\frac{1}{u_1z^2-z+1}$.
Suppose there exists some singularity $\zeta\in\mathbb{C}$ which is a root of
$u_1z^2-z+1$. By construction $\zeta\neq 0$ and $\zeta$ is necessarily a
singularity of $W(z)$.
Suppose $\vert \zeta\vert \le \kappa_k$, then we arrive at the contradiction
$\vert W(\zeta)\vert > W(\kappa_k)$
since $W(\zeta)$ is not finite and $W(\kappa_k)=
\sum_{n\ge 0}f_k(2n,0)\rho_k^{2n}<\infty$.
Therefore all dominant singularities of $\sum_{n \ge 0}{\sf C}_{k}^{}(n)\, z^n$
are dominant singularities of $W(z)$. By Pringsheim's
Theorem~\cite{Titmarsh:39},
$\sum_{n \ge 0}{\sf C}_{k}^{}(n)\, z^n$ has a dominant positive real
singularity which by construction equals $\rho_k$ and the Claim follows. \\
The Claim immediately implies that the exponential growth rate is the
inverse of the minimal positive real solution of the equation
$\frac{\sqrt{u_1}\ x}{u_1x^2-x+1}  =  \rho_k$.
According to \cite{Wang:07} the power series
$\sum_{n\ge 0}f_k(2n,0)z^n$ has an analytic continuation in a
$\Delta_{\rho_k}$-domain and we have
$[z^n]W(z)\sim K \varphi_k(n) (\rho^{-1})^n$, where $\varphi_k(n)$ is given by
eq.~(\ref{E:f-k-imp}).
We can therefore employ Theorem~\ref{T:transfer}, which via eq.~(\ref{E:C1})
allows us to transfer the subexponential factors from the asymptotic
expressions for $f_k(2n,0)$ to ${\sf C}_{k}(n)$. From this
eq.~(\ref{E:growth-free}) follows and the proof of Theorem~\ref{T:core}
is complete.
\end{proof}

%%%
%%%%%%%%%%%%%%%%%%%%%%%%%%%%%%%%%%%%%%%%%%%%%%%%%%%%%%%%%%%%%%%%%%%%%%%%%
%%%

\section{Pseudoknot RNA with stack-length $\ge \sigma$}\label{S:arc-2}

%%%
%%%%%%%%%%%%%%%%%%%%%%%%%%%%%%%%%%%%%%%%%%%%%%%%%%%%%%%%%%%%%%%%%%%%%%%%%
%%%

In this Section we combine Lemma~\ref{L:arc-2} and Lemma~\ref{L:www} in order
to derive the generating function of $k$-noncrossing RNA pseudoknot structures
with minimum stack-size $\sigma$. It is worth mentioning that core-structures
are only implicit (via Lemma~\ref{L:www}) in its proof: all expressions and
relations are based on ${\sf T}_{k,1}(n',h')$ and ${\sf T}_{k,1}(n)$,
respectively. The latter are given by Theorem~\ref{T:cool1}.
Our main result reads
%%%
%%%%%%%%%%%%%%%%%%%%%%%%%%%%%%%%%%%%%%%%%%%%%%%%%%%%%%%%%%%%%%%%%%%%%%
%%%
\begin{theorem}\label{T:arc-2}
Let $k,\sigma\in\mathbb{N}$, $k\ge 2$, let $x$ be an indeterminant and
$\rho_k$ the dominant, positive real singularity of
$\sum_{n\ge 0}f_k(2n,0)z^n$ (eq.~(\ref{E:rho-k})).
Then
\begin{eqnarray*}
{\sf T}_{k,\sigma}^{}(n,h) & = & \sum_{b=\sigma-1}^{h-1}\sum_{j=0}^{(h-b)-1}
{b+(2-\sigma)(h-b)-1 \choose h-b-1}(-1)^{(h-b)-j-1} \\
 & & \qquad \qquad \qquad \times \, {(h-b)-1 \choose j}
{\sf T}^{ }_{k,1}(n-2h+2j+2,j+1) \ .
\end{eqnarray*}
Furthermore, ${\sf T}^{}_{k,\sigma}(n)$,
satisfies the following identity
\begin{equation}\label{E:f1}
\sum_{n \ge 0}{\sf T}_{k,\sigma}^{}(n)x^n =
\frac{1}{u_0x^2-x+1}\sum_{n \ge
0}f_{k}(2n,0)\left(\frac{\sqrt{u_0}x}{u_0x^2-x+1}\right)^{2n} \ ,
\end{equation}
where $u_0=\frac{(x^2)^{\sigma-1}}{(x^2)^{\sigma}-x^2+1}$.
Furthermore
\begin{equation}\label{E:growth-2}
{\sf T}_{k,\sigma}^{}(n) \sim \varphi_k(n)\,
\left(\frac{1}{\gamma_{k,\sigma}^{}}\right)^n
\end{equation}
holds, where $\gamma_{k,\sigma}^{}$ is a dominant singularity of
$\sum_{n \ge 0}{\sf T}_{k,\sigma}^{}(n)x^n$ and the minimal positive real
solution of the equation
\begin{equation}
\frac{\sqrt{\frac{(x^2)^{\sigma-1}}{(x^2)^{\sigma}-x^2+1}}\ x}
{\left(\frac{(x^2)^{\sigma-1}}{(x^2)^{\sigma}-x^2+1}\right)\,
x^2-x+1}=\rho_k
\end{equation}
and $\varphi_k(n)$ is a polynomial over $n$ derived from the
asymptotic expression of $f_k(2n,0)\sim \varphi_k(n)\left(\frac{1}
{\rho_k}\right)^n$ of eq.~(\ref{E:f-k-imp}).
\end{theorem}
%%%%
%%%%%%%%%%%%%%%%%%%%%%%%%%%%%%%%%%%%%%%%%%%%%%%%%%%%%%%%
%%%%

In the following we present the first $18$ numbers of ${\sf T}_{3,2}(n)$,
${\sf T}_{3,3}(n)$, ${\sf T}_{4,2}(n)$ and ${\sf T}_{4,3}(n)$:
\begin{center}
\begin{tabular}{c|cccccccccccccccccc}
%\hline
%  \multicolumn{9}{|c|}{\bf $k$-noncrossing RNA}\\
$n$ & \small $1$ & \small $2$ & \small $3$ &\small $4$ & \small $5$
& \small $6$ & \small $7$ & \small$8$ & \small$9$ & \small$10$ &
\small $11$ & \small $12$ & \small $13$ &\small $14$ & \small $15$ &
\small $16$ & \small $17$ & \small$18$ \\
\hline${\sf T}_{3,2}(n)$ & \small $1$ & \small $1$ & \small$1$ &
\small $1$  & \small $2$ & \small$4$ & \small$8$ & \small $15$ &
\small $28$ & \small $55$ & \small $110$ & \small $222$ &\small
$448$ &\small $913$
&\small $1890$ &\small $3964$ &\small $8385$ &\small $17846$\\
${\sf T}_{3,3}(n)$ & \small $1$ & \small $1$ & \small$1$ & \small
$1$ & \small $1$ & \small$1$ & \small$2$ & \small $4$ & \small $8$ &
\small $14$ & \small $23$ & \small$36$ & \small $56$ & \small $91$ &
\small$155$ & \small$275$ & \small $491$
& \small $869$\\
\end{tabular}
\end{center}

\begin{center}
\begin{tabular}{c|cccccccccccccccccc}
%\hline
%  \multicolumn{9}{|c|}{\bf $k$-noncrossing RNA}\\
$n$ & \small $1$ & \small $2$ & \small $ 3$ &\small $4$ & \small $5$
& \small $6$ & \small $7$ & \small $8$ & \small$9$ & \small$10$ &
\small $11$ & \small $12$ & \small $13$ &\small $14$ & \small $15$ &
\small $16$ & \small $17$
& \small $18$ \\
\hline${\sf T}_{4,2}(n)$ & \small $1$ & \small $1$ & \small$1$ &
\small $1$  & \small $2$ & \small$4$ & \small$8$ & \small $15$ &
\small $28$ & \small $55$ & \small $110$ & \small $223$ &\small
$455$ &\small $944$
&\small $1995$ &\small $4274$ &\small $9244$ &\small $20182$\\
${\sf T}_{4,3}(n)$ & \small $1$ & \small $1$ & \small$1$ & \small
$1$ & \small $1$ & \small$1$ & \small$2$ & \small $4$ & \small $8$ &
\small $14$ & \small $23$ & \small$36$ & \small $56$ & \small $91$ &
\small$155$ & \small$275$ & \small $491$
& \small $870$\\
\end{tabular}
\end{center}

\begin{proof}
The first assertion follows from Lemma~\ref{L:core} and eq.~(\ref{E:Moebius}),
which allows to express the terms ${\sf C}_k(n-2b,h-b)$ via
${\sf T}_{k,1}(n',h')$.
In order to prove eq.~(\ref{E:growth-2})
we apply Lemma~\ref{L:www} twice. First,
Lemma~\ref{L:www} implies for arbitrary $\sigma$ and $u=1$
\begin{equation}
\sum_{n \ge 0}{\sf T}_{k,\sigma}^{ }(n)x^n = \sum_{n \ge
0}\sum_{h \le \frac{n}{2}}{\sf
C}_{k}^{ }(n,h)\left(\frac{(x^2)^{\sigma-1}}{1-x^2}\right)^hx^n
+ \frac{x}{1-x}
\end{equation}
and secondly, it guarantees for arbitrary $u\in \mathbb{C}$ and
$\sigma=1$
\begin{equation}\label{E:whow}
\sum_{n \ge 0}\sum_{h\le \frac{n}{2}}{\sf
T}_{k,1}^{ }(n,h)u^hx^n = \sum_{n \ge 0}\sum_{h \le
\frac{n}{2}}{\sf
C}_{k}^{ }(n,h)\left(\frac{u}{1-ux^2}\right)^hx^n + \frac{x}{1-x}\ .
\end{equation}
The key observation (the ``bridge'') is here the relation between
$\sigma$ and $u$ via the terms $\frac{(x^2)^{\sigma-1}}{1-x^2}$ and
$\frac{u}{1-ux^2}$. It is clear that for any $\sigma\in\mathbb{N}$
there exists an unique solution $u_0$ for
\begin{equation}\label{E:key}
\frac{(x^2)^{\sigma-1}}{1-x^2} =\frac{u}{1-ux^2}
\end{equation}
given by $u_0=\frac{(x^2)^{\sigma-1}}{(x^2)^{\sigma}-x^2+1}$. This
allows to express
$$
\sum_{n \ge
0}\sum_{h \le \frac{n}{2}}{\sf
C}_{k}^{ }(n,h)\left(\frac{(x^2)^{\sigma-1}}{1-x^2}\right)^hx^n
+ \frac{x}{1-x}
$$
for any $\sigma$ via the bivariate generating function
$\sum_{n \ge 0} \sum_{h\le \frac{n}{2}}{\sf
T}_{k,1}^{}(n,h)\ u^hx^n$. Now we employ
Lemma~\ref{L:arc-2}, which provides an interpretation of the latter
as follows:
\begin{equation}
\sum_{n \ge 0}\sum_{h\le \frac{n}{2}}{\sf T}_{k,1}^{}(n,h)u^hx^n=
\frac{1}{ux^2-x+1}\sum_{n \ge
0}f_{k}(2n,0)\left(\frac{\sqrt{u}x}{ux^2-x+1}\right)^{2n} \ .
\end{equation}
We accordingly obtain
\begin{eqnarray*}
\sum_{n \ge 0}{\sf T}_{k,\sigma}^{}(n)x^n & = & \sum_{n \ge 0}
\sum_{h \le \frac{n}{2}}{\sf C}_{k}^{}(n,h)\left(
\frac{(x^2)^{\sigma-1}}{1-x^2}\right)^hx^n + \frac{x}{1-x}\\
& = & \sum_{n \ge 0}\sum_{h \le \frac{n}{2}}{\sf
C}_{k}^{}(n,h)\left(\frac{u_0}{1-u_0x^2}\right)^hx^n + \frac{x}{1-x}\\
 & = & \frac{1}{u_0x^2-x+1}\sum_{n \ge
0}f_{k}(2n,0)\left(\frac{\sqrt{u_0}x}{u_0x^2-x+1}\right)^{2n} \ .
\end{eqnarray*}
and eq.~(\ref{E:f1}) follows.
We set $V(z)=\sum_{n \ge 0}f_{k}(2n,0)\left(\frac{\sqrt{u_0} z}
{u_0z^2-z+1}\right)^{2n}$.\\
%%%%%%%%%%%%%%%%%%%%%%%%%%%%%%%%%%%%%%%%%%%%%%%%%%%%%%%%%%%%%%%%%%%%%%%%%%
{\it Claim.} All dominant singularities of $\sum_{n\ge 0}
            {\sf T}_{k,\sigma}^{}(n) z^n$ are singularities
            of $V(z)$ and in particular $\gamma_{k,\sigma}$ is a dominant
            singularity. \\
%%%%%%%%%%%%%%%%%%%%%%%%%%%%%%%%%%%%%%%%%%%%%%%%%%%%%%%%%%%%%%%%%%%%%%%%%
To prove the Claim we observe that a dominant singularity of
$$
 \frac{1}{u_0x^2-x+1}\sum_{n \ge
0}f_{k}(2n,0)\left(\frac{\sqrt{u_0}x}{u_0x^2-x+1}\right)^{2n}
$$
is either a singularity of $V(z)$ or $\frac{1}{u_0z^2-z+1}$.
Suppose there exists some singularity $\zeta\in\mathbb{C}$ which is a root of
$\frac{1}{u_0z^2-z+1}$. By construction $\zeta\neq 0$ and $\zeta$ is
necessarily a singularity of $V(z)$. Suppose $\vert \zeta\vert \le \kappa_k$,
then we arrive at the contradiction
$\vert V(\zeta)\vert>\vert V(\kappa_k)\vert$ since $V(\zeta)$
is not finite and
$$
V(\kappa_k)=\sum_{n\ge 0}f_k(2n,0)\rho_k^{2n}<\infty \ .
$$
Therefore all dominant singularities of
$\sum_{n\ge 0} {\sf T}_{k,\sigma}^{}(n) z^n$
are singularities of $V(z)$. By Pringsheim's
Theorem~\cite{Titmarsh:39},  $\sum_{n\ge 0}
            {\sf T}_{k,\sigma}^{}(n) z^n$ has a dominant positive real
singularity which by construction equals $\gamma_{k,\sigma}$ and the
Claim follows.\\
The equation
\begin{equation}
\frac{\sqrt{\frac{(x^2)^{\sigma-1}}{(x^2)^{\sigma}-x^2+1}}\ x}
{\left(\frac{(x^2)^{\sigma-1}}{(x^2)^{\sigma}-x^2+1}\right)\,
x^2-x+1}=\rho_k
\end{equation}
has a minimal positive real solution and the Claim implies that its inverse
equals the exponential growth-rate.
According to \cite{Wang:07} the power series
$\sum_{n\ge 0}f_k(2n,0)z^n$ has an analytic continuation in a
$\Delta_{\rho_k}$-domain and we have
$[z^n]V(z)\sim K \varphi_k(n) (\rho^{-1})^n$, where $\varphi_k(n)$ is given by
eq.~(\ref{E:f-k-imp}).
In view of eq.~(\ref{E:f1}) we can therefore
employ Theorem~\ref{T:transfer}, which allows us to transfer the
subexponential factors from the asymptotic expressions for $f_k(2n,0)$ to
${\sf T}_{k,\sigma}(n)$, whence eq.~(\ref{E:growth-2}).
This completes the proof of
Theorem~\ref{T:arc-2}.
\end{proof}

%%%
%%%%%%%%%%%%%%%%%%%%%%%%%%%%%%%%%%%%%%%%%%%%%%%%%%%%%%%%%%%%%%%%%%%%%%%%%%
%%%
{\bf Acknowledgments.}
%%%
%%%%%%%%%%%%%%%%%%%%%%%%%%%%%%%%%%%%%%%%%%%%%%%%%%%%%%%%%%%%%%%%%%%%%%%%%%
%%%
We are grateful to Prof.~William Y.C.~Chen for helpful comments. Special thanks to
Jing Qin for creating Figures~\ref{F:lego0} and \ref{F:lego2}.
This work was supported by the 973 Project, the PCSIRT Project of the
Ministry of Education, the Ministry of Science and Technology, and
the National Science Foundation of China.

\bibliographystyle{plain}
\bibliographystyle{plain}

%%%
%%%%%%%%%%%%%%%%%%%%%%%%%%%%%%%%%%%%%%%%%%%%%%%%%%%%%%%%%%%%%%%%%%%%%%%%%%
%%%

\end{document}